\documentclass{appolb}
\usepackage{graphicx}
\begin{document}
% \eqsec  % uncomment this line to get equations numbered by (sec.num)
\title{Study of the Onset of Deconfinement and Search for the\\
Critical Point of Strongly Interacting Matter\\
at the CERN SPS %
\thanks{Presented at 42nd Int. Symposium on Multiparticle Dynamics, Kielce, Poland}%
}
\author{Peter Seyboth
\address{Max-Planck-Institut f\"ur Physik, Munich \\
 and Jan Kochanowski University, Kielce}
}
\maketitle
\begin{abstract}
Collisions of lead nuclei have been studied at the CERN SPS since 1994.
A review is presented of the evidence for the production of deconfined matter, 
the location of the energy of the onset of deconfinement and the search
for the critical point of stronly interacting matter. 
\end{abstract}
  
\section{Introduction}

\begin{figure}[htb]
\centerline{%
\includegraphics[width=5.0cm]{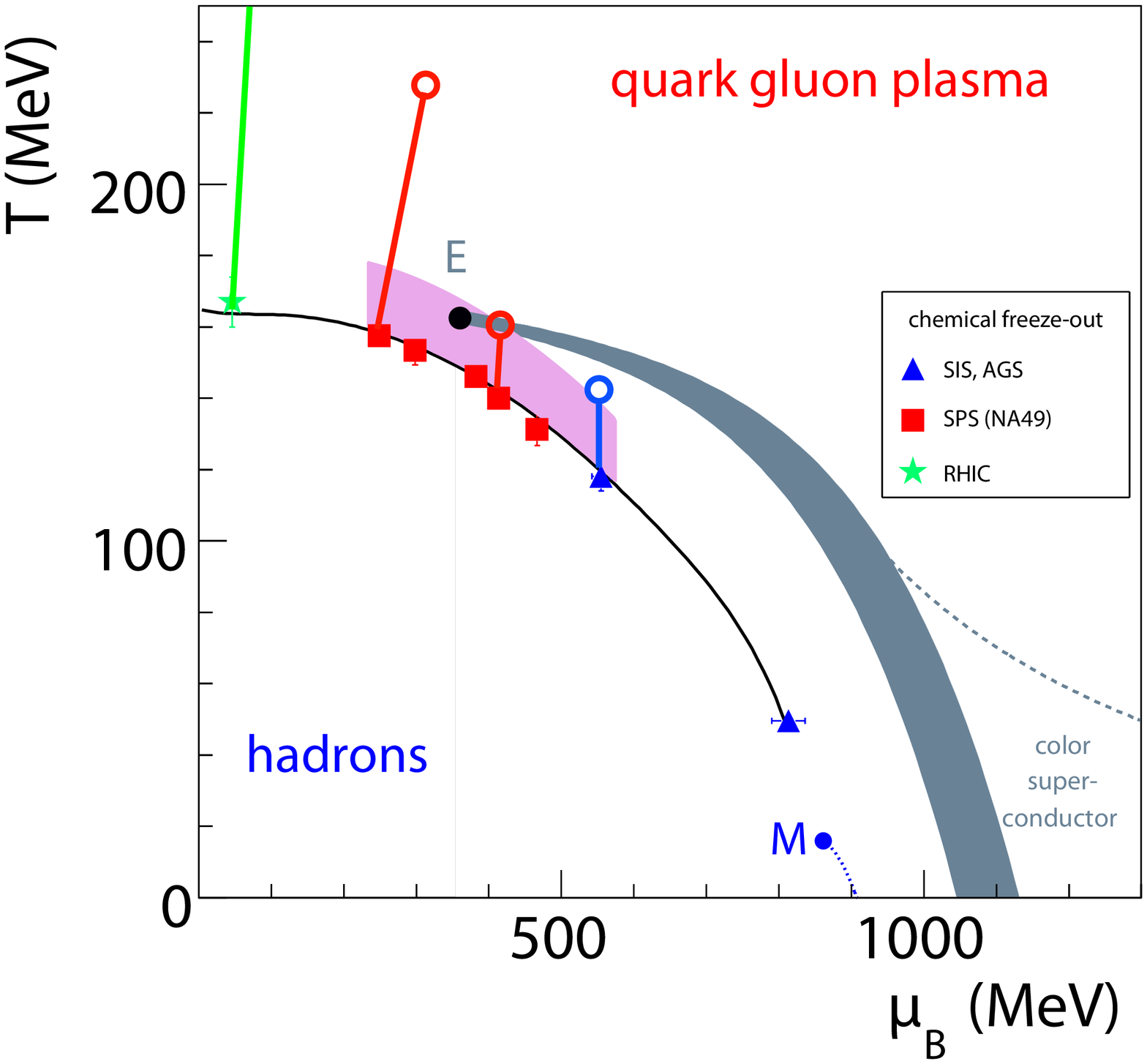}
\includegraphics[width=6.0cm]{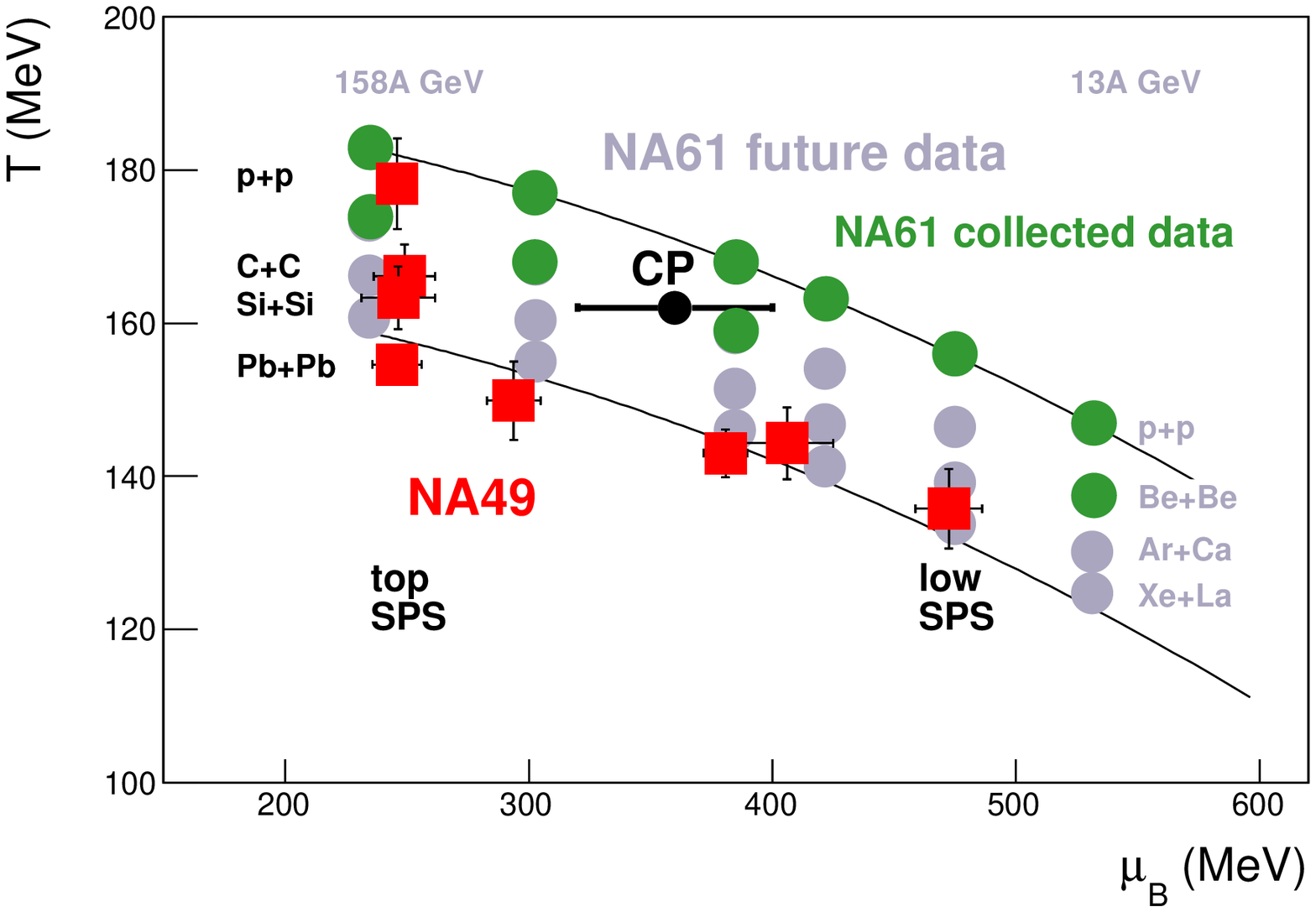}}
\caption{Left: Phase diagram of strongly interacting matter. Locations of  
freezeout points of hadron composition were obtained from statistical model 
fits to particle yields. Right: reactions studied by experiments NA49 (squares) 
and NA61 (green dots -- recorded, grey dots -- planned).}
\label{FIG:phase_diagram}
\end{figure}

The success of the quark model of hadrons and the discovery of point-like partons
in nucleons and scaling of the cross sections in deep-inelastic scattering experiments 
led to the development of quantum chromodynamics QCD. Ordinary hadron
matter is composed of hadrons in which the constituent quarks and gluons are
confined. It was soon conjectured that at high temperature and/or pressure
hadrons would dissolve into quasi-free quarks and gluons \cite{collins_perry_1975}
the quark gluon plasma (QGP). The development of QCD calculations on the lattice
then allowed to explore the non-perturbative aspects of the theory, namely the
phase diagram of strongly interacting matter. As schematically shown in 
Fig.~\ref{FIG:phase_diagram}~left, the existence region of hadrons at low temperature T and 
baryochemical potential $\mu_{B}$ is thought to be separated from the QGP phase
at high T by a first order phase boundary (shaded band), which ends in a critical point E
and then turns into a crossover at low  $\mu_{B}$. The experimental investigation of
the phase diagram and of the properties of the QGP is the main purpose of the study
of heavy-ion collisions.

Pb+Pb reactions were investigated at the CERN SPS since 1994 by a variety of
experiments in the available energy range from 158$A$ down to 20$A$~GeV
($\sqrt{s_{NN}} = $ 17.3 -- 6.3 GeV). At top energy an initial energy density of 
$\approx$~3~GeV/fm$^3$~\cite{na49_eps} was found, sufficient for deconfinement.
Many of the predicted signatures for the QGP were observed~\cite{heinz_jacob_2000},
e.g. enhancement of strangeness production due to the smaller mass of
strangeness carriers (s~quark versus $K$~meson), suppression of charmonium
production in excess of cold nuclear matter effects due to screening of the
color force, indication of photon emission
from annihilation of the copious quarks and antiquarks in the QGP and 
strong modification of the $\rho^0$~meson line shape possibly connected to
chiral restoration. However, none of these signatures turned out to be
uniquely attributable to the QGP.

The NA49 collaboration therefore proposed an energy scan (T and $\mu_{B}$ at freezeout
increases respectively decreases with collisions energy) in order to look
for a sharp change in hadron production properties which were predicted
to signal the onset of deconfinement at the early stage of evolution of
the collision fireball~\cite{gazdzicki_gorenstein_1999}. In fact,
these signals were observed by NA49~\cite{na49_alt_2008} and their system size
dependence is under study by the successor expeiment NA61~\cite{na61_proposal}. 
The data from the energy scan also
allowed to start the search for the predicted critical point CP. First 
tantalising hints obtained by NA49 are presently pursued by experiment NA61~\cite{na61_proposal} 
with a two dimensional scan of the phase diagram in nuclear size A and 
energy, by which the freezeout point is moved in T and $\mu_{B}$ 
(see Fig.~\ref{FIG:phase_diagram}(right)). The expected signature of 
the CP is a maximum of fluctuations in the produced final state when 
freezeout occurs close to the CP location~\cite{stephanov_al_1999}.

\section{Onset of deconfinement}

The results discussed in this section were mostly obtained by NA49 
which was the only experiment to participate in the entire energy scan 
at the SPS (1999 -- 2002). Measurements are thus restricted to charged hadrons 
detected in the large acceptance NA49 apparatus~\cite{na49_det_1999} 
consisting of time projection chambers (TPC) as the principal detectors.
Results (see Ref.~\cite{na49_alt_2008}) are compared to statistical model 
predictions and interpreations of the data, in particular the SMES 
model~\cite{gazdzicki_gorenstein_1999} which motivated the SPS energy scan.
 
\begin{figure}[htb]
\centerline{%
\includegraphics[width=6.0cm]{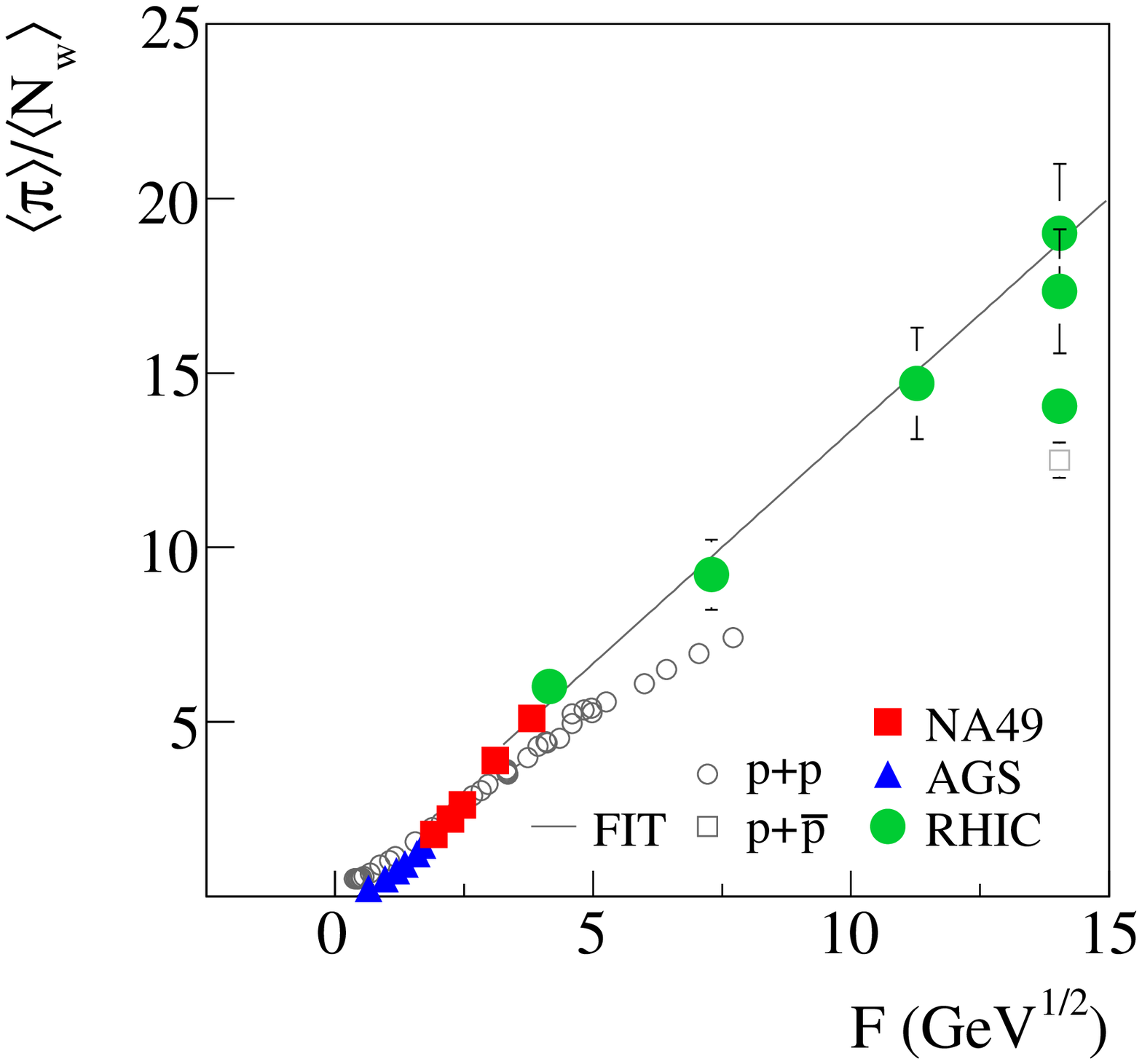}
\includegraphics[width=6.0cm]{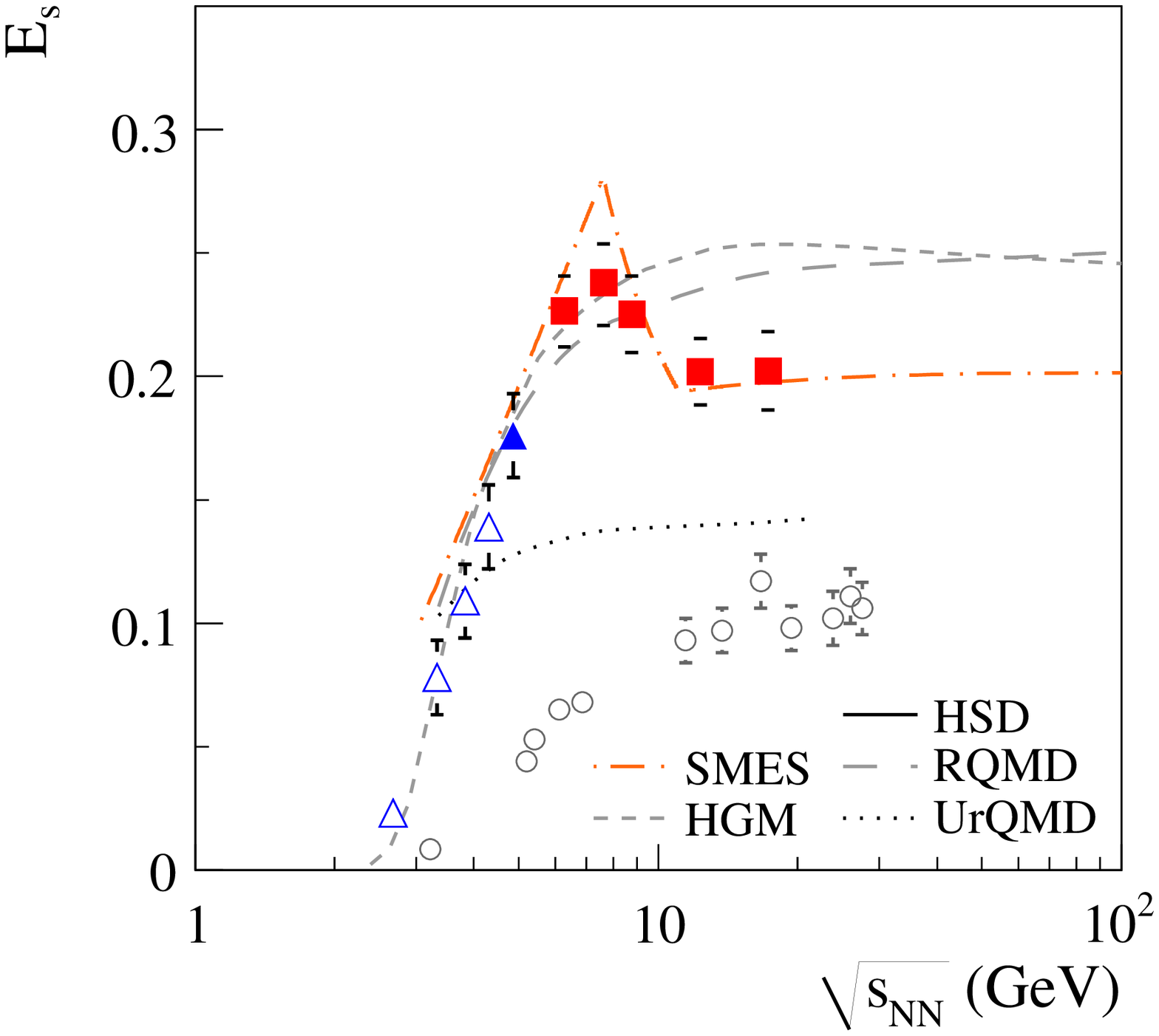}}
\caption{Left: mean pion multiplicity $\langle \pi \rangle$ per wounded 
nucleon $\langle N_W \rangle$  measured in central Pb+Pb and Au+Au
collisions (full symbols) and $p(\bar{p})+p$ collisons (open symbols)
versus Fermi energy variable F~$\approx s_{NN}^{0.25}$.
Right: ratio E$_s$ (see text) versus collision energy $\sqrt{s_{NN}}$
compared to model calculations.}
\label{FIG:pi2nw_es}
\end{figure}

As seen in Fig.~\ref{FIG:pi2nw_es}~(left) the pion yield per wounded nucleon
starts to increase faster in central Pb+Pb and Au+Au collisions than in $p+p(\bar{p})$
reactions at the low end of the SPS energy range. Pion production is a measure
of entropy production and the steepening of slope indicates a 3-fold increase
of the initial degrees of freedom as expected for deconfinement. Figure~\ref{FIG:pi2nw_es}
displays the energy dependence of the ratio of strangeness to pion yields which
is given to good accuracy by $E_s = (\langle K \rangle + \langle \Lambda \rangle )/ \langle \pi \rangle $
at SPS energies. One observes a threshold rise to a sharp peak (the horn) and a decrease to a
plateau value. In the SMES these features were predicted as arising from the 
increase of temperature in the hadron phase, the onset of deconfinement at the
peak and the plateau (with predicted height) when all of the fireball is in the QGP phase 
initially.

\begin{figure}[htb]
\centerline{%
\includegraphics[width=6.0cm]{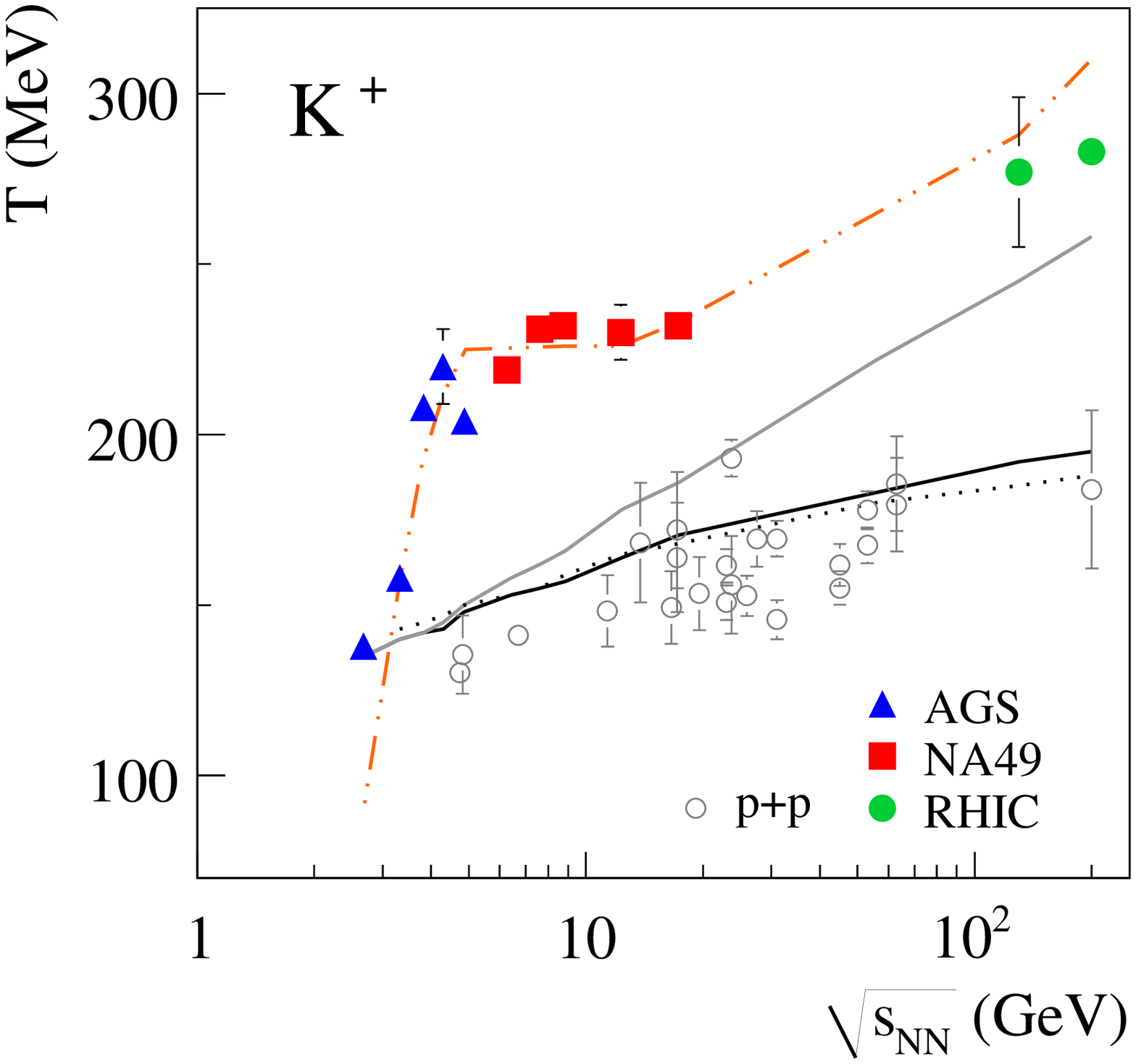}
\includegraphics[width=5.0cm]{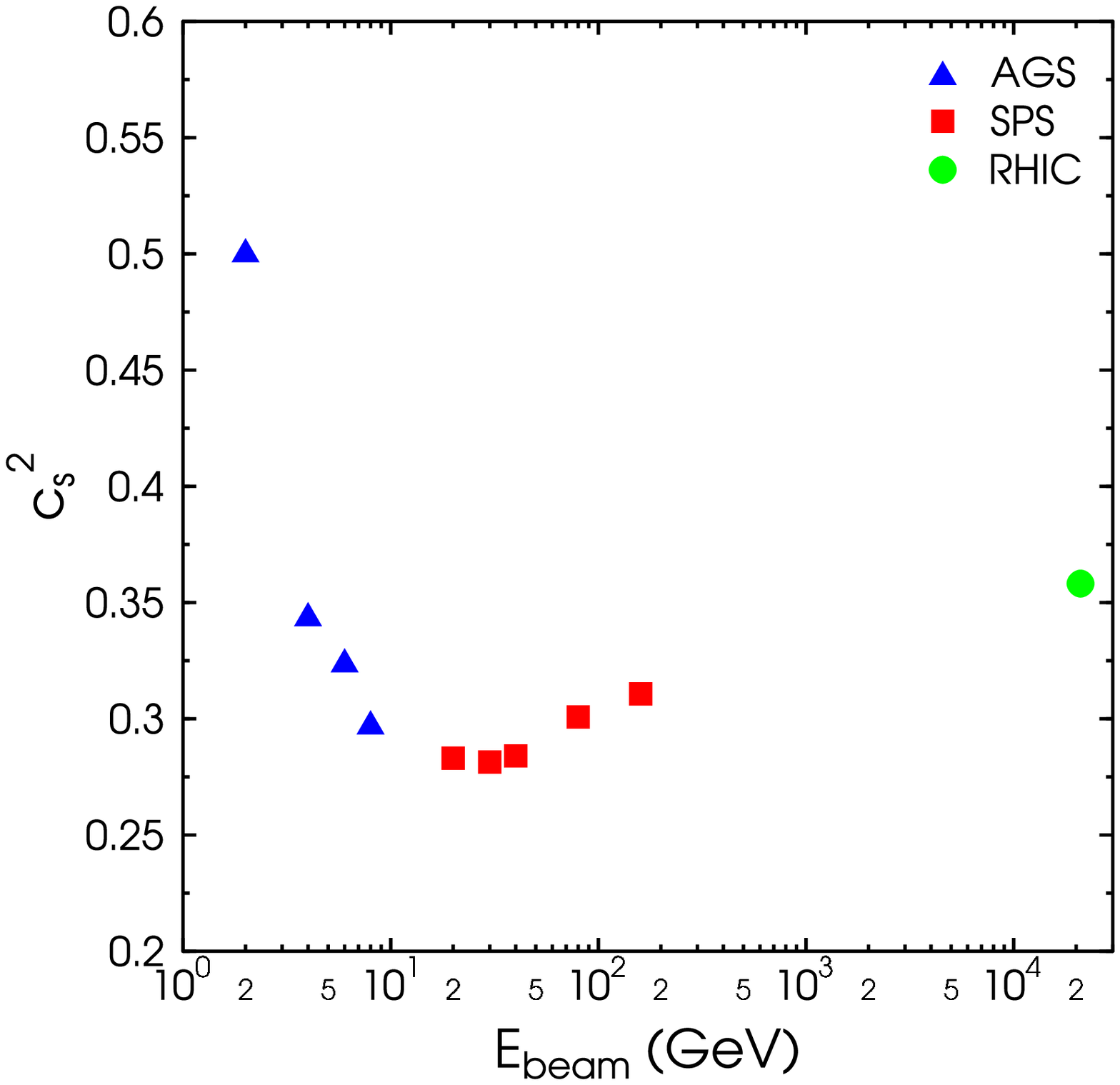}}
\caption{Left: inverse slope paramter T of $K^+$ transverse mass spectra
compared to model calculations. Filled (open) symbols show results from
central Pb+Pb and Au+Au (p+p) collisions. 
Right: velocity of sound $c_s$ derived from the width of pion rapidity
spectra.}
\label{FIG:tkpls_cs}
\end{figure}

Onset of deconfinement is expected to be accompanied by a soft region of the equation
of state EoS~\cite{softpoint} leading to observable structures in the energy dependence 
of transverse and longitudinal particle spectra. After a threshold rise, a stationary value 
(step) is found in the SPS region for the inverse slope parameter T for $K^+$ 
(see~Fig.~\ref{FIG:tkpls_cs}~(left)) which characterises the temperature and the 
transverse flow in the produced fireball~\cite{step_2003}.
Similar behaviour is observed for $K^-$ mesons, $\pi$ mesons and protons. The sound 
velocity $c_s$ in the fireball can be deduced from the width of the rapidity
distribution of pions~\cite{bleicher_2006}. The results, plotted in 
Fig.~\ref{FIG:tkpls_cs}~(right), show a minimum at SPS energies as expected
for a soft mixed phase at the onset of deconfinement.

\begin{figure}[htb]
\centerline{%
\includegraphics[width=6.0cm]{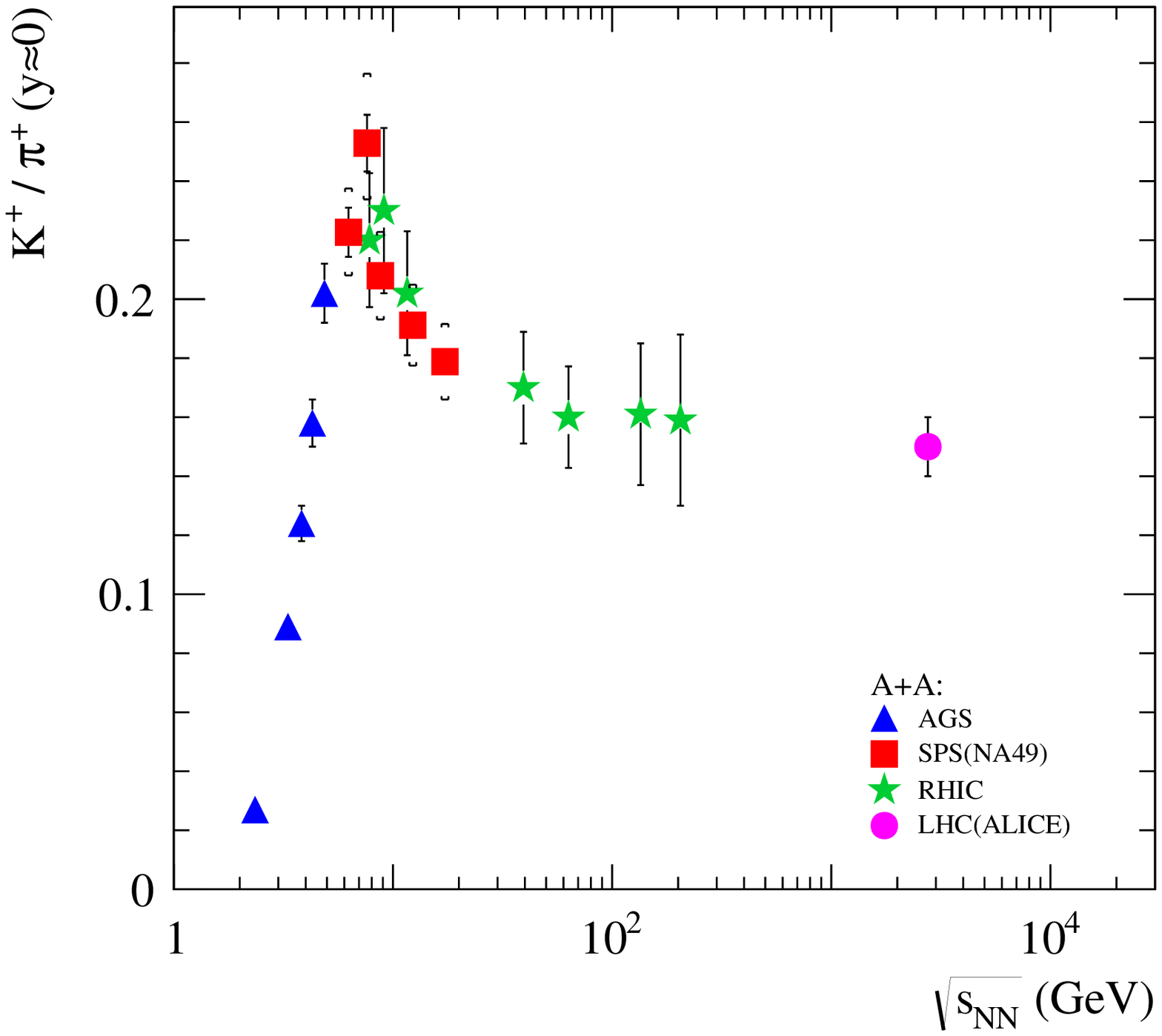}
\includegraphics[width=6.0cm]{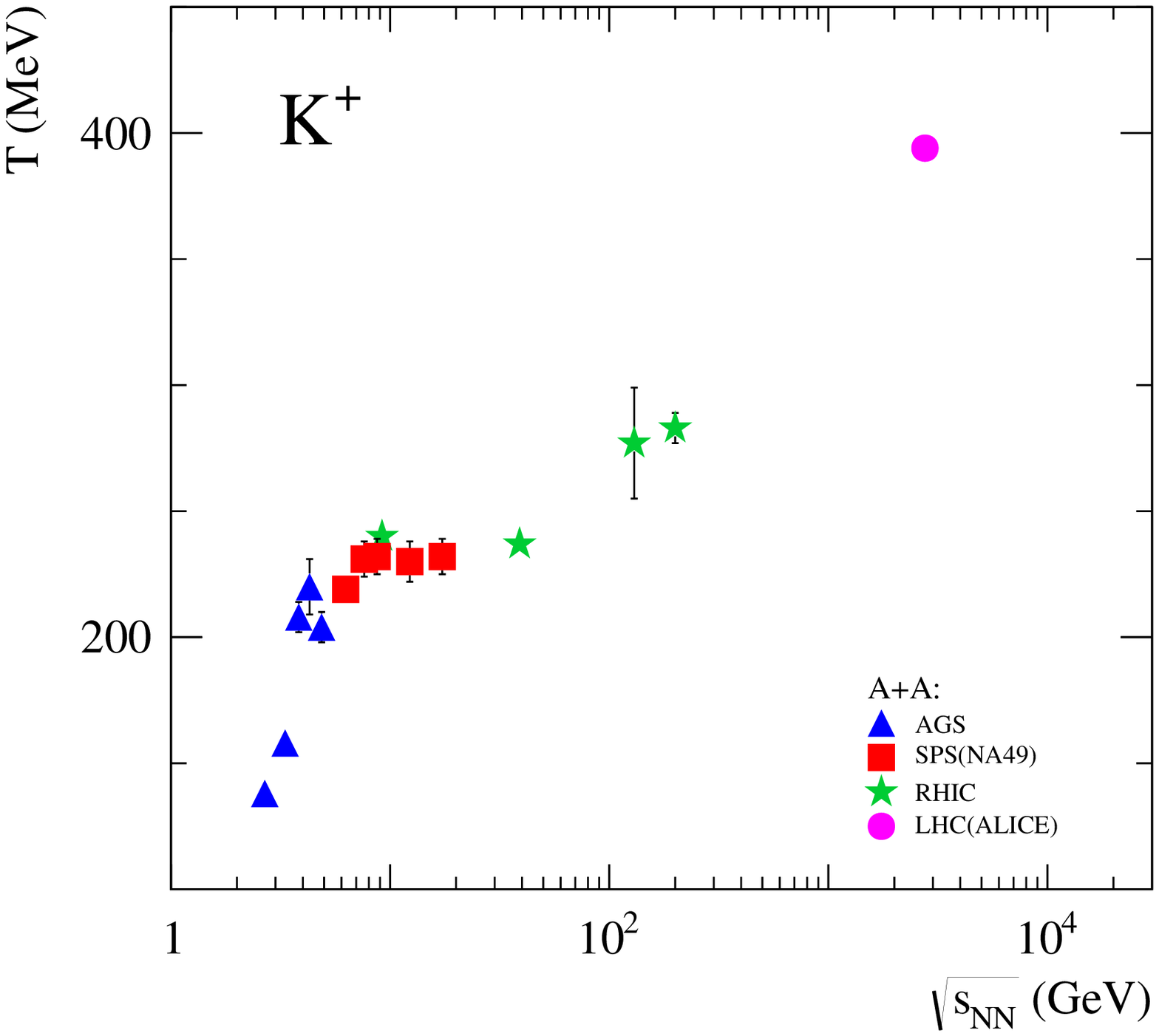}}
\caption{Left: yield ratio $K^+/\pi^+$ at midrapidity.  
Right: inverse slope parameter T of $K^+$ transverse mass spectra.
Plotted data on central Pb+Pb and Au+Au collisions are from E802 at AGS (triangles), 
NA49 at SPS (squares), STAR at RHIC (stars) and ALICE at LHC (dot).}
\label{FIG:bes_horn_step}
\end{figure}

The discussed features, observed in central Pb+Pb and Au+Au collisions,
are only explained by models which incorporate a phase transition for
energies above about 30$A$~GeV (see dash-dotted curves in 
Figs.~\ref{FIG:pi2nw_es},\ref{FIG:tkpls_cs}). Thus the most natural explanation
of the hadron production measurements is the onset of deconfinement at low SPS 
energies~\cite{ggs_2011}. Recent results of STAR from the RHIC low energy 
scan confirm the horn and step structures found by NA49 and measurements 
of ALICE at the LHC agree with the expected trends towards 
highest energy~(see~Fig.~\ref{FIG:bes_horn_step}). 

\begin{figure}[htb]
\centerline{%
\includegraphics[width=4.0cm]{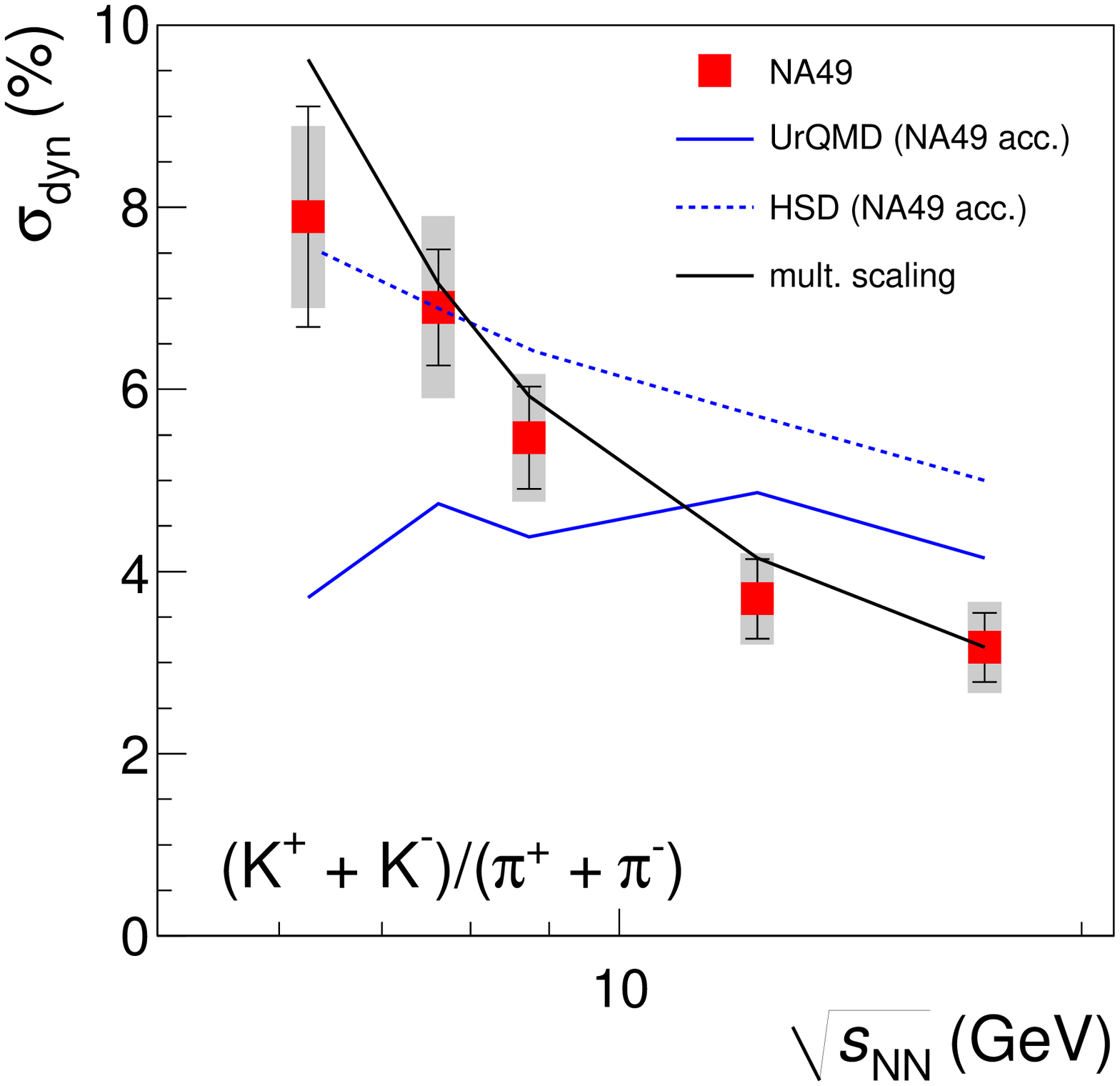}
\includegraphics[width=4.0cm]{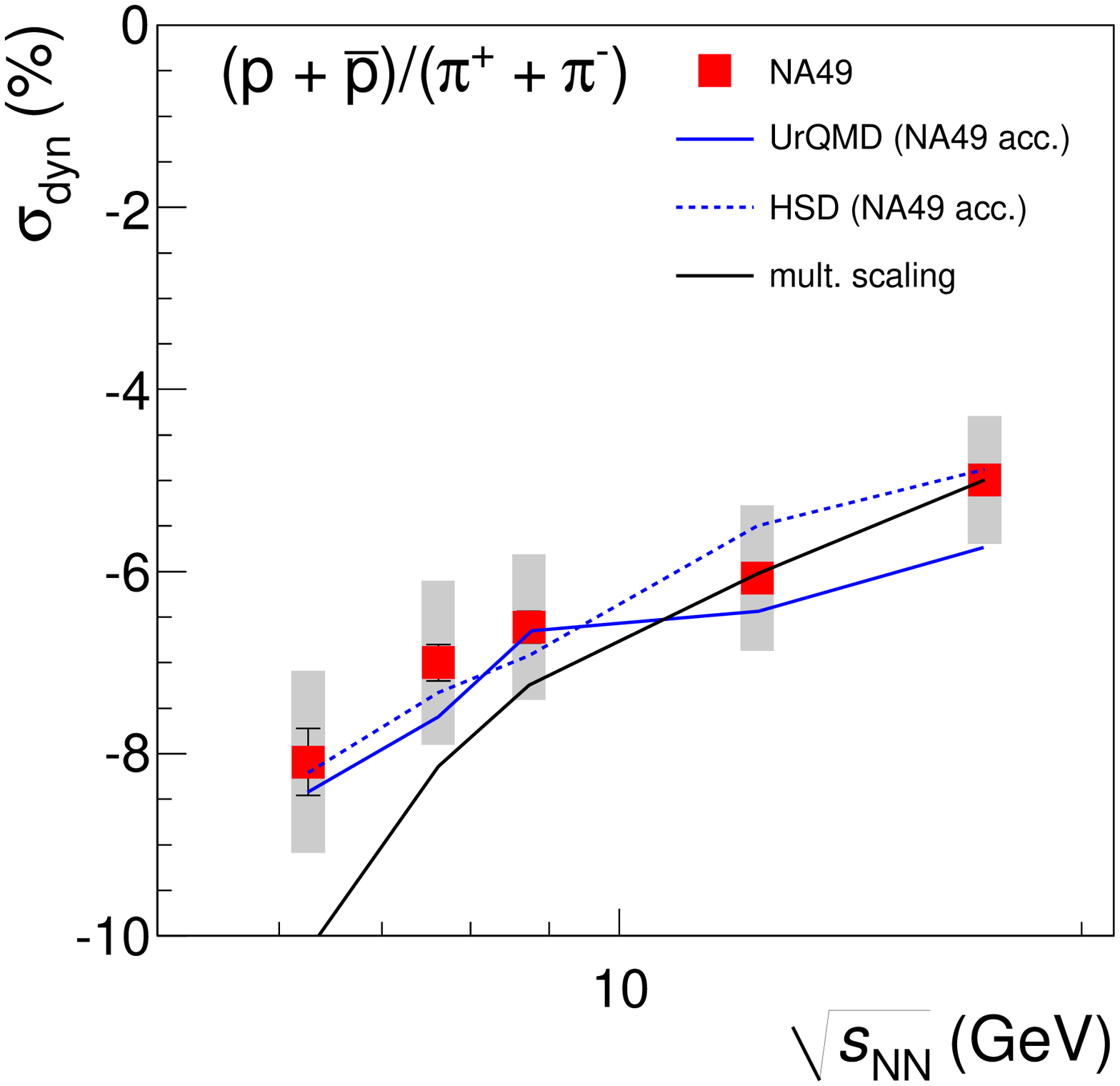}
\includegraphics[width=4.0cm]{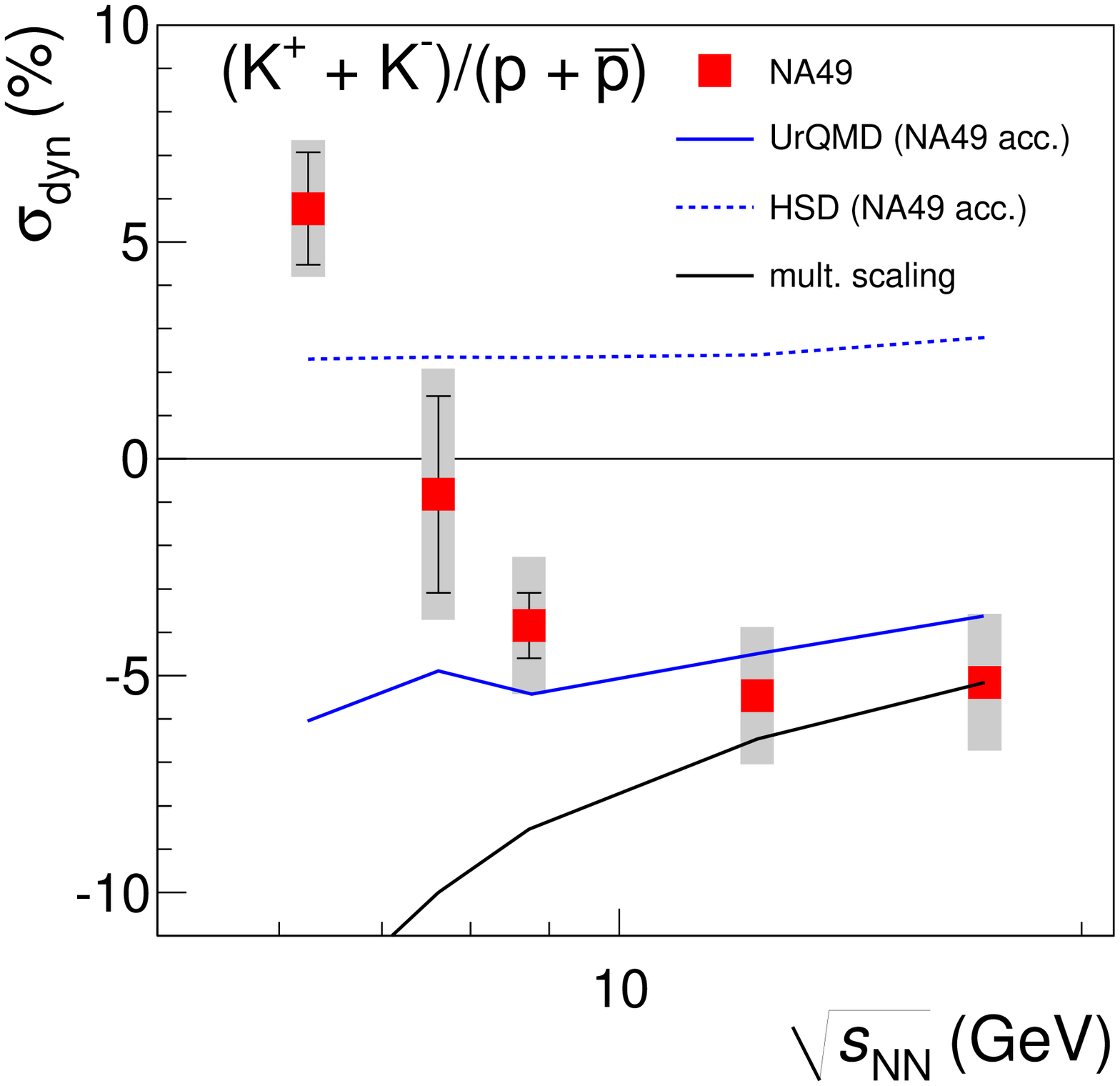}}
\caption{Event-by-event fluctuations of particle ratios K/$\pi$~(left), 
p/$\pi$~(center) and K/p~(right) versus collision energy in central
Pb+Pb collisions~\cite{na49_ratio_fluct} expressed in terms of the
observable $\sigma_{dyn}$. Curves show comparisons to
a multiplicity scaling model~\cite{rfluct_multscaling}, 
and microscopic models UrQMD and HSD.} 
\label{FIG:ratio_fluct}
\end{figure}

Crossing of a phase transition during the evolution of the produced particle
system may lead to a modification of event-by-event fluctuations of various 
quantities. In particular, the different ratio of strange to non-strange degrees
of freedom in hadron matter and QGP might give rise to fluctuations of the 
K/$\pi$ ratio~(Fig.~\ref{FIG:ratio_fluct}~(left)) near the onset of
deconfinement~\cite{go_ga_zo_fluct_2004}. A rise towards lowest SPS energy is 
observed~\cite{na49_ratio_fluct}, but it is well described by 
generic multiplicity scaling~\cite{rfluct_multscaling}
of the studied observable $\sigma_{dyn}$ assuming constant intrinsic 
correlations. While the same conclusion holds for the p/$\pi$ ratio, 
there is a sign change for the fluctuations of K/p. This indicates a qualitative 
change of correlations between baryons and strange particles~\cite{koch_sbcorr_2005}.
Recent measurements of the related measure $\nu_{dyn}$ by STAR~\cite{STAR_fluct} 
did not show any energy dependence. This difference has now been traced to the 
different phase space acceptance region covered by NA49 and STAR~\cite{na49_identity}.

The smaller charge of quarks in the QGP compared to hadrons was expected to
reduce fluctuations of the net charge observed in not too large regions of
phase space~\cite{cfluct}. The effect has not been confirmed by measurements~\cite{na49_cfluct}
and is believed to be erased by the hadronisation process. In case of a phase
transition the evolution time of the fireball until freezeout is expected
to increase and become manifest in a widening of the charge balance function~\cite{qbalance}. 
Although such behaviour was found for Pb+Pb reactions at the highest SPS energy for
increasingly central collisions~\cite{na49_qbalance}, the effect may also be
caused by increasing radial flow and local charge conservation.

\section{Search for the critical point of strongly interacting matter}

\begin{figure}[htb]
\centerline{%
\includegraphics[width=4.0cm]{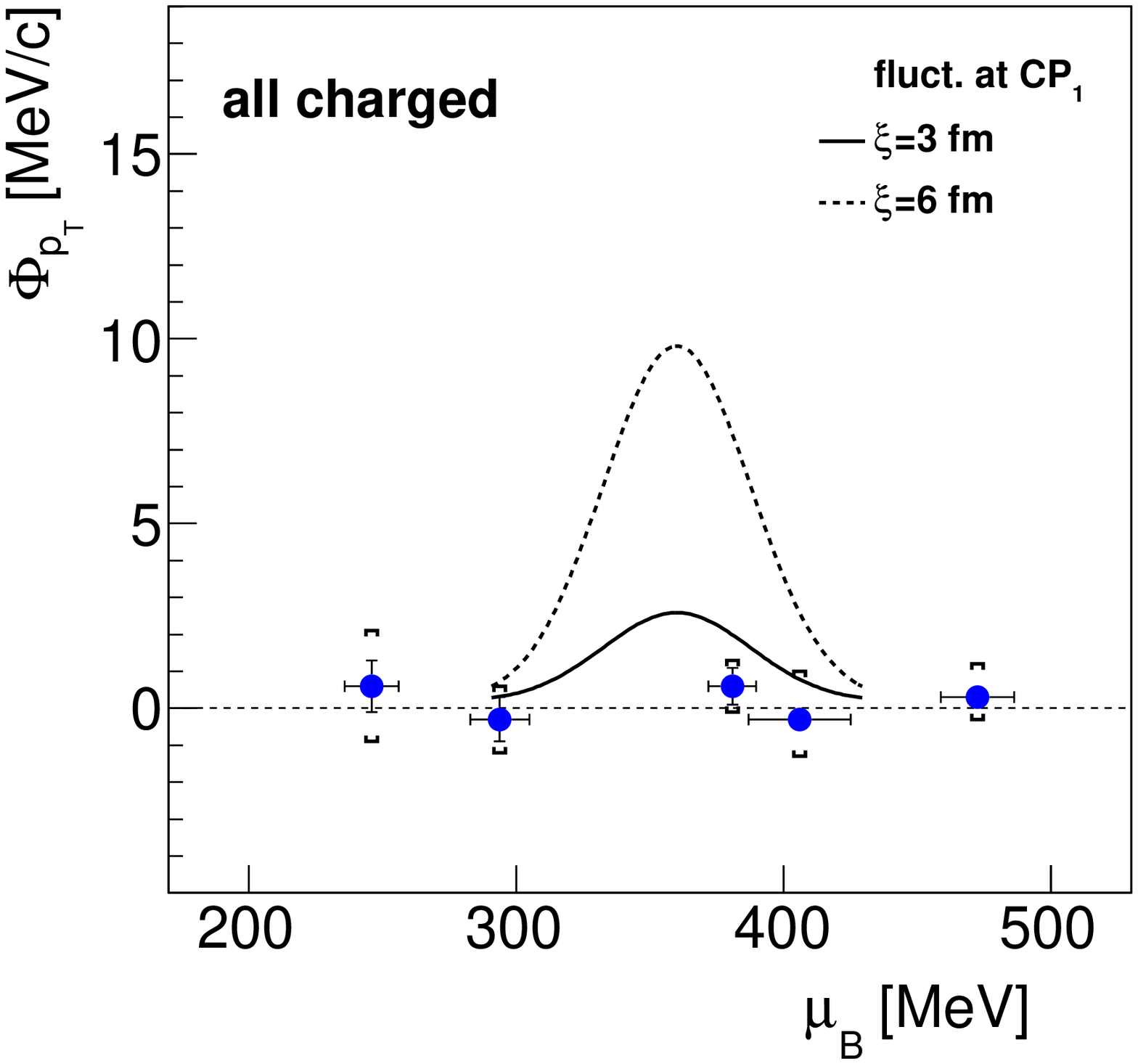}
\includegraphics[width=4.0cm]{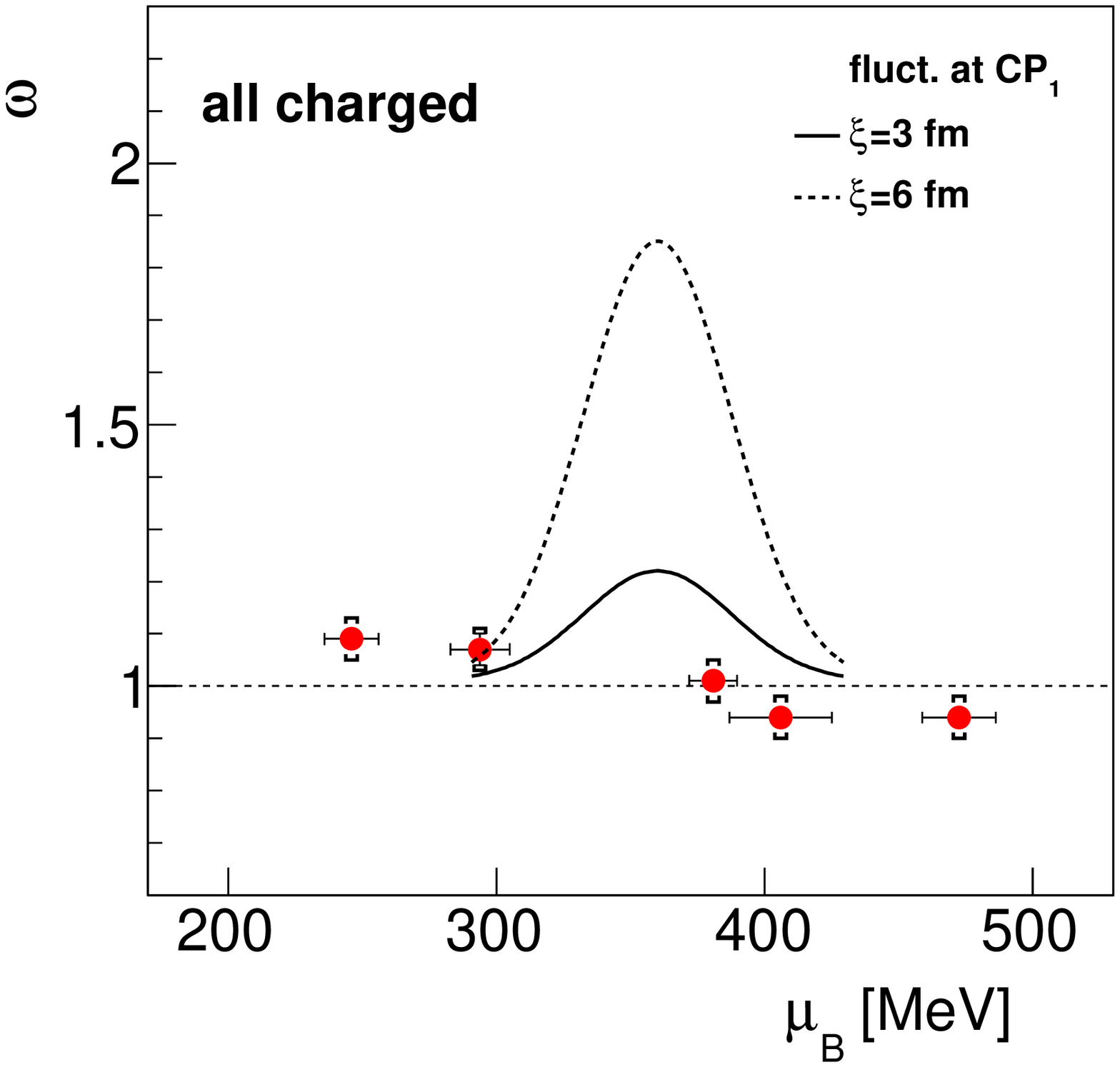}
\includegraphics[width=4.0cm]{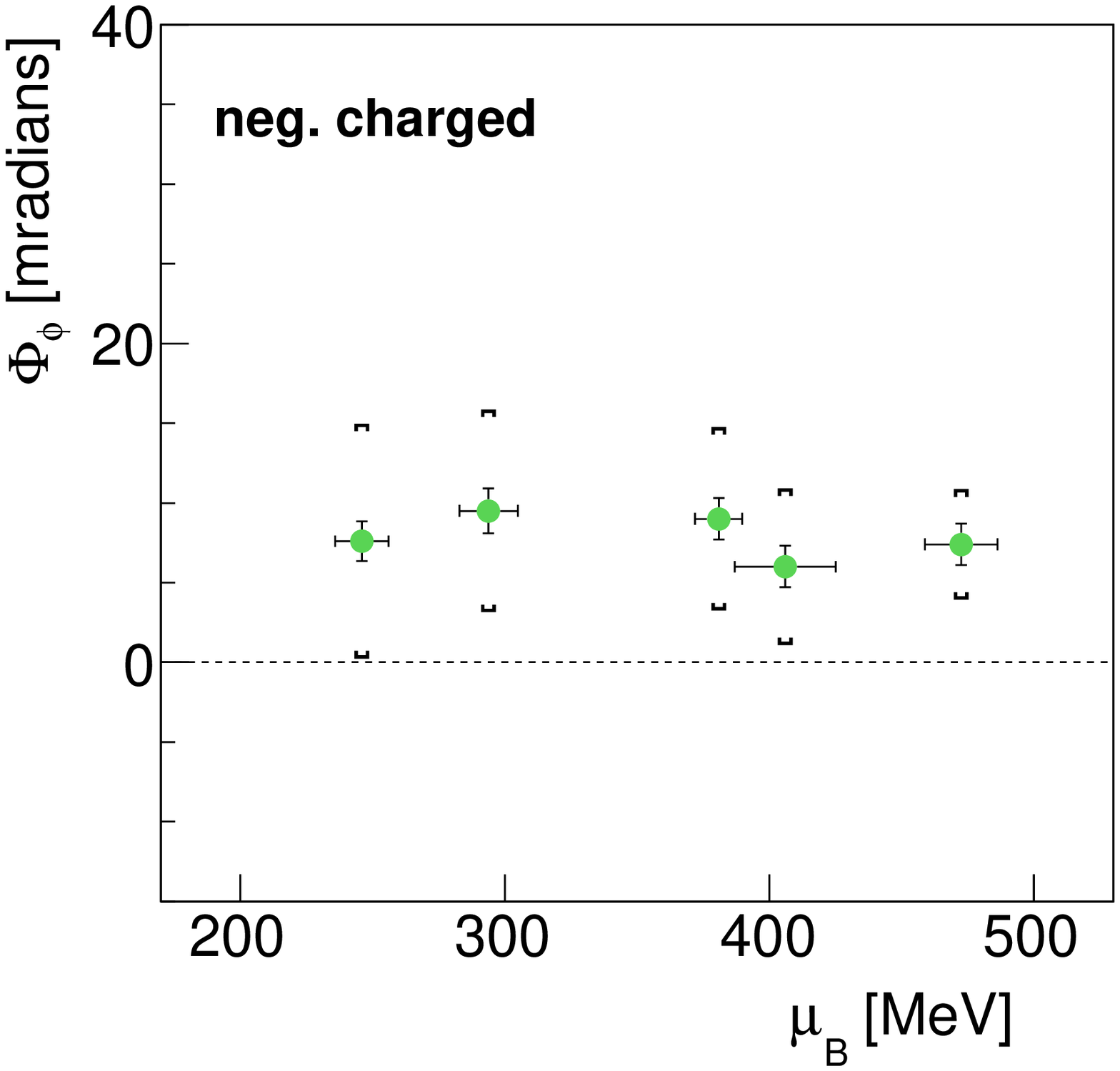}}
\vspace{-1.0cm}
\end{figure}
\begin{figure}[htb]
\centerline{%
\includegraphics[width=4.0cm]{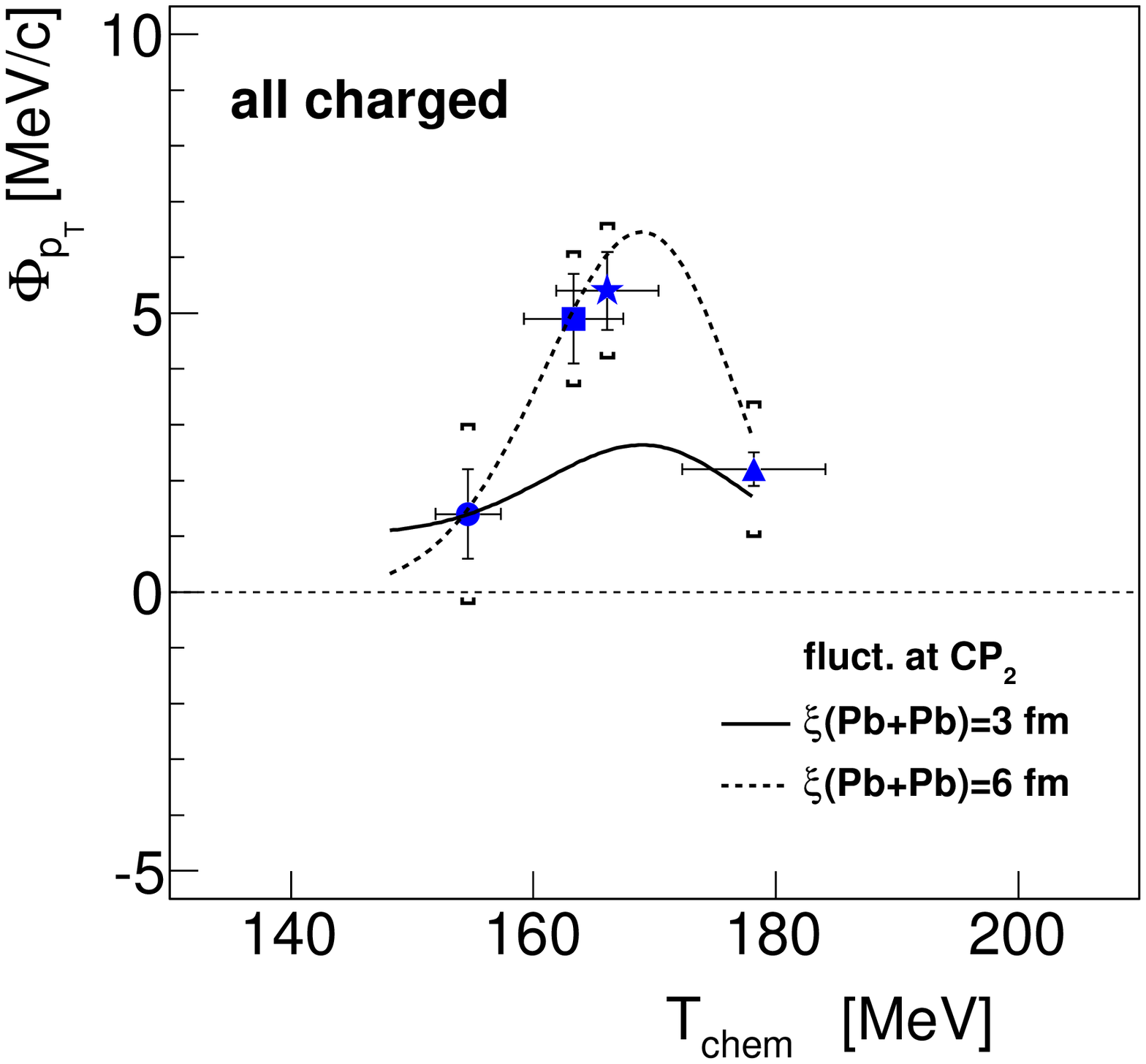}
\includegraphics[width=4.0cm]{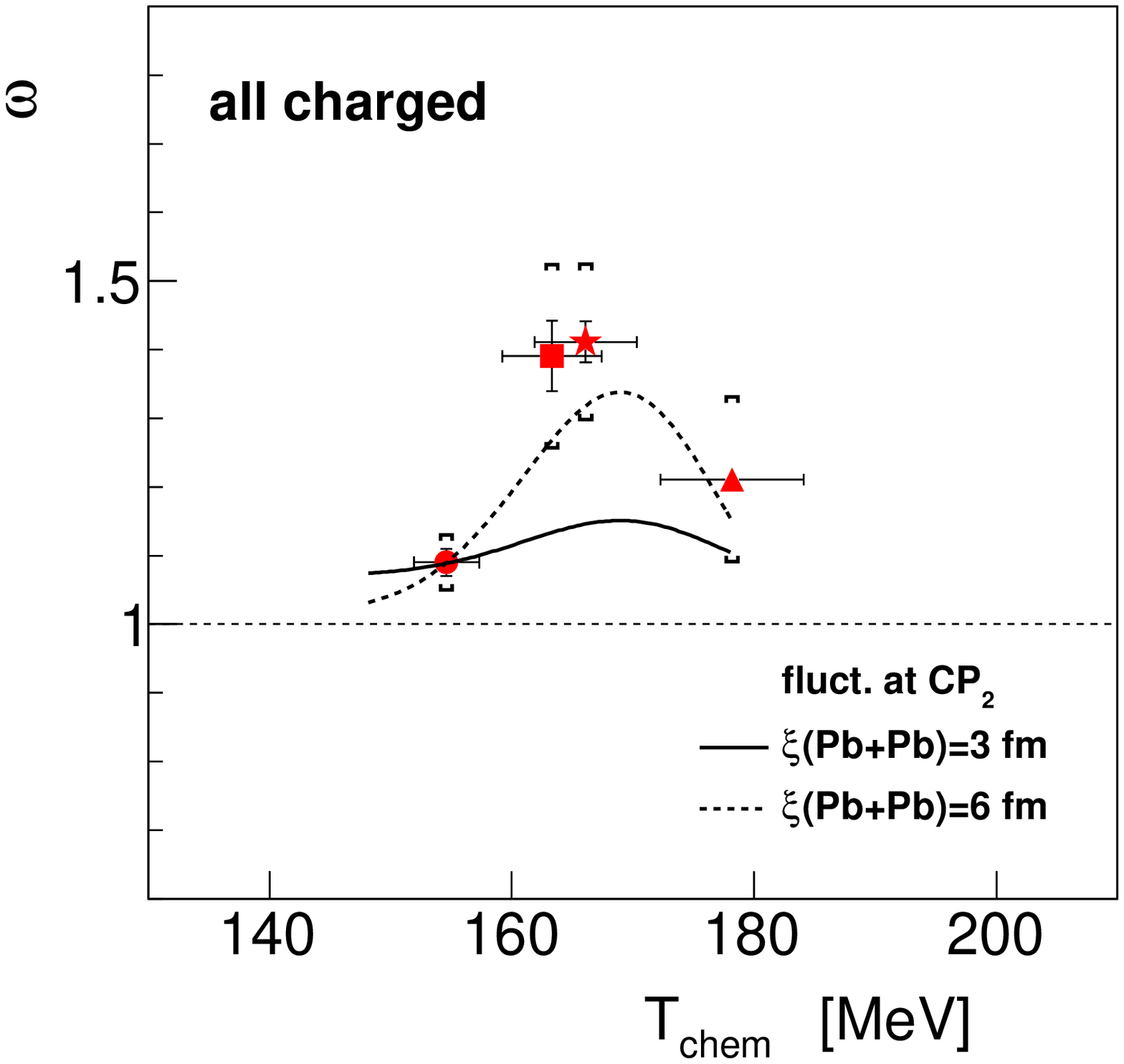}
\includegraphics[width=4.0cm]{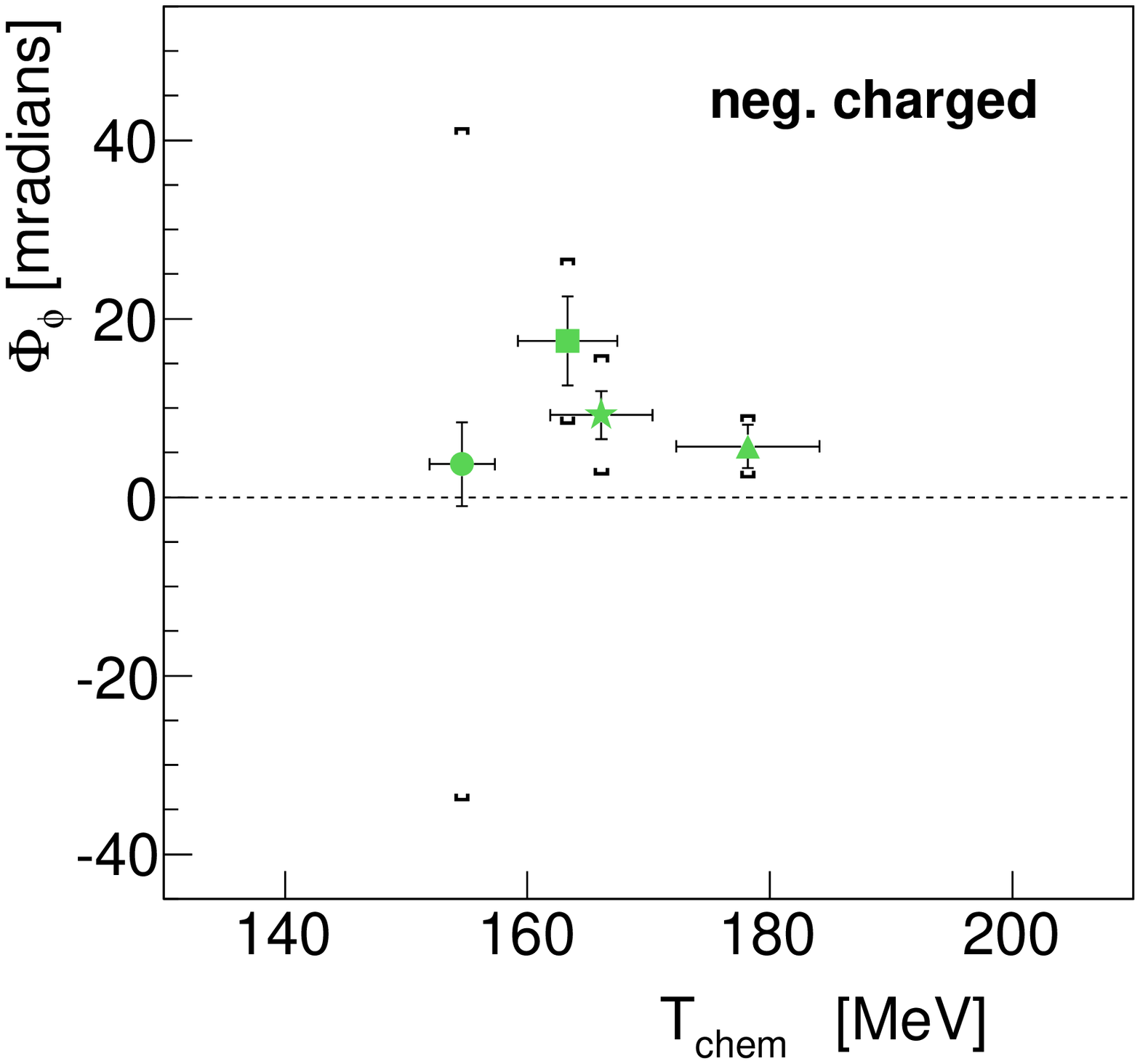}}
\caption{Fluctuation measures $\Phi_{p_T}$ for $\langle p_T \rangle$ (left~\cite{na49_fipt_omega_qm2009}),
scaled variance $\omega$ of the charged particle multiplicity distribution (center~\cite{na49_fipt_omega_qm2009})
and $\Phi_{\phi}$ for the average azimuthal angle $\langle \phi \rangle$ (right~(NA49 preliminary))
versus baryochemical potential $\mu_B$ (top row) and temperature T (bottom row) in cms rapidity $1.1 < y < 2.6$.
Values of $\mu_B$ and T were estimated from statistical model fits~\cite{stat_model}
to particle yields.}
\label{FIG:na49_fipt_omega_fifi}
\end{figure}

Lattice QCD calculations for finite $\mu_B$ are difficult and are still
under development. Nevertheless, several groups obtained quantitative
estimates~\cite{cpoint_lqcd} placing the CP in the region of the
phase diagram which is accessible at SPS energies. If the evolution trajectory
of the fireball passes close enough to the critical point enhanced fluctuations
from event to event are predicted~\cite{stephanov_al_1999}. The size of such fluctuations
is limited by the lifetime and the size of the fireball (correlation length
less than 3--6 fm in central Pb+Pb collisions) and they may be
erased by rescattering processes in the late hadron stage. In spite of such caveats
an intense search for the CP and its fluctuation signatures in e.g.
average transverse momentum $\langle p_T \rangle$, multiplicities and
local density fluctuations is in progress.

Measurements were performed for the energy scan of central Pb+Pb collisions as well 
as for central C+C, Si+Si and inelastic p+p reactions at highest SPS energy of 158$A$~GeV.
Results are plotted in Fig.~\ref{FIG:na49_fipt_omega_fifi} as function of the phase diagram
variables T and $\mu_B$ of the fireball freezeout as obtained from statistical model
fits~\cite{stat_model}. The fluctuation measures $\Phi_{p_T}$, $\omega$ and $\Phi_{\phi}$
vary smoothly with $\mu_B$ (top row in Fig.~\ref{FIG:na49_fipt_omega_fifi}). However, there
is an indication of a maximum in T for the lighter collision systems (see bottom row in
Fig.~\ref{FIG:na49_fipt_omega_fifi}). Curves show estimates of the possible effect of a 
CP~\cite{na49_fipt_omega_qm2009,stephanov_pc}.

\begin{figure}[htb]
\centerline{%
\includegraphics[width=4.0cm]{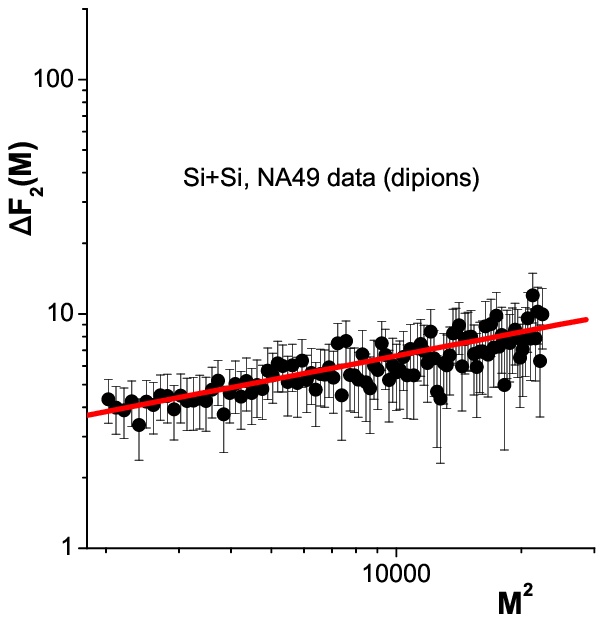}
\includegraphics[width=4.0cm]{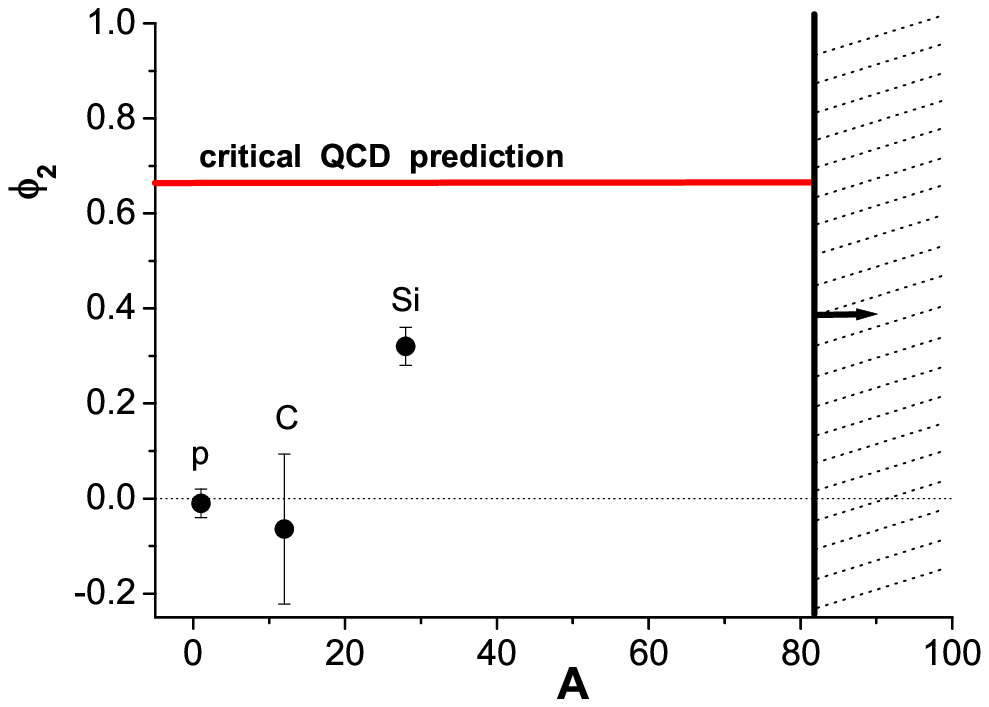}
\includegraphics[width=4.0cm]{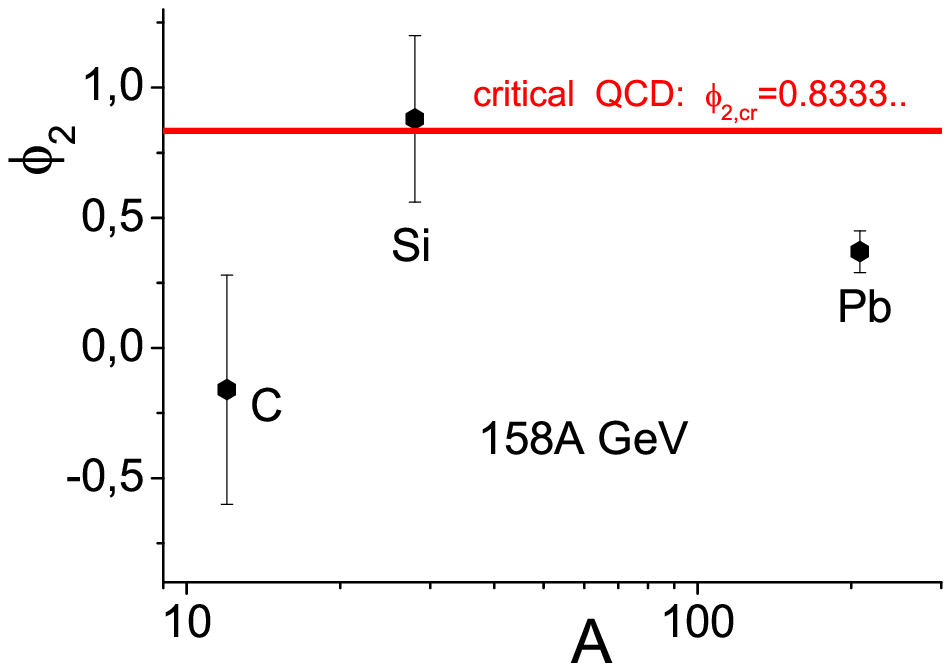}}
\caption{Left: background corrected scaled factorial moment $\Delta F_2$ versus number of cells $M^2$ in
transverse momentum space for central Si+Si collisions at 158$A$~GeV~\cite{na49_int_pipi}.
Center and right: fitted intermittency index $\Phi_2$ for low-mass $\pi^+\pi^-$ pairs~\cite{na49_int_pipi}
and protons~\cite{na49_int_prot} in p+p and central C+C, Si+Si and Pb+Pb collisions.}
\label{FIG:na49_intermittency}
\end{figure}

Another promising search strategy for the CP is the study of local density fluctuations
of low-mass $\pi^+\pi^-$ pairs and protons produced near midrapidity. Theoretical 
investigations~\cite{crit_fluct} expect power law fluctuations near the CP which can be studied by the
intermittency analysis method in transverse momentum space using scaled factorial moments $F_n(M)$
where M is the number of subdivisions in each $p_T$ direction. 
The power law exponents from fits of the form $F_2(M) \propto M^{2\Phi_2}$ can be predicted
and should be approached near the CP. In the experimental analysis 
fits are applied to $\Delta F_2(M)$, the background
corrected scaled factorial moment (example with superimposed fit is
shown in Fig.~\ref{FIG:na49_intermittency}~(left)). Results for $\Phi_2$ are plotted in
Fig.~\ref{FIG:na49_intermittency}~(center,~left) for low-mass $\pi^+\pi^-$ pairs and protons,
respectively. As found for the integrated fluctuation observables in the preceding paragraph, 
there is also an indication of a maximum and an approach to the CP prediction for the 
local density fluctuations in the Si+Si system. These tantalising hints need further 
confirmation and strongly motivate the more precise and systematic studies under way in NA61.

\section{Conclusion}

Evidence for the onset of deconfinement was found in central Pb+Pb collisions
at the CERN SPS and confirmed by by the RHIC beam energy scan program.
The critical point of strongly interacting matter has so far eluded
convincing detection in collisions of heavy nuclei. The most suitable
energy region for continued search is the range of the SPS energies which is
being intensively studied by NA61, the beam energy scan BES at RHIC and 
in future also by NICA in Dubna. The ultra-high energies of the LHC,
on the other hand, offer  the best environment for measuring the properties 
of the QGP. The planned program at SIS-100 at FAIR in Darmstadt covers the lowest
energy, high baryon density domain for which other exotic forms of matter
are predicted.

%uncomment the following lines to place a figure
%\begin{figure}[htb]
%\centerline{%
%\includegraphics[width=12.5cm]{Fig1}}
%\caption{Plot of ...}
%\label{Fig:F2H}
%\end{figure}

\end{document}